\documentclass[11pt]{article}

\usepackage{cite}
\usepackage{amsfonts}
\usepackage{amsmath}
\usepackage{latexsym}
\usepackage{fancybox}
\usepackage[dvipdfmx]{graphicx}
\usepackage[dvipdfmx]{color}

\numberwithin{equation}{section}
\allowdisplaybreaks

\def\spa#1{\phantom{\fbox{\rule[-#1cm]{0cm}{0cm}}}}
\DeclareMathOperator{\arccosh}{arccosh}

\def\[#1\]{\begin{align}#1\end{align}}
\newcommand\dashint{\mathchoice
  {\int\kern-10pt-}
  {\int\kern-8.5pt-}
  {\int\kern-6.1pt-}
  {\int\kern-4.58pt-}}
\def\vev#1{\left\langle#1\right\rangle}
\begin{document}

\hfuzz=100pt
\title{{\Large \bf{
Properties of the wormhole-dominant phase \\in two-dimensional quantum gravity
}}}
%\date{}
\author{Tsunehide Kuroki$^{a}$\footnote{kuroki@toyota-ti.ac.jp} and Yuki Sato$^{b,c}$\footnote{yukisato@u-fukui.ac.jp} 
  \spa{0.5} \\
\\
$^a${\small{\it Theoretical Physics Laboratory, Toyota Technological Institute}}
\\ {\small{\it 2-12-1 Hisakata, Tempaku-ku, Nagoya 468-8511, Japan}}\\    
\\
$^b${\small{\it Department of Mechanical and System Engineering, University of Fukui}}
\\ {\small{\it 3-9-1 Bunkyo, Fukui-shi, Fukui 910-8507, Japan}}\\    
\\
$^c${\small{\it Department of Physics, Nagoya University}}
\\ {\small{\it Chikusaku, Nagoya 464-8602, Japan}}\\  
\spa{0.3} 
}
\date{}

\maketitle
\centerline{}

\begin{abstract} 
We study the $N \times N$ Hermitian one-matrix model modified by the double-trace interaction. 
It is known that the coupling for the double-trace interaction can control the weight for the microscopic wormholes 
if interpreting the matrix model as the lattice model of random surface; 
tuning the coupling to its critical value, the effect of wormholes become substantial to change the critical behavior of the pure $2$D quantum gravity, 
which is characterized by a certain positive value of the string susceptibility.    

In the large-$N$ limit, we calculate the continuum limit of the disk amplitude in which the wormhole effects are important. 
The resulting continuum disk amplitude is the same as that of the pure $2$D quantum gravity.  
We also introduce the renormalized coupling for the double-trace interaction, 
and show that the newly introduced renormalized coupling can alter the renormalized bulk cosmological constant effectively.

\end{abstract}

\renewcommand{\thefootnote}{\arabic{footnote}}
\setcounter{footnote}{2}

\newpage

%%%%%%%%%%%%%%%%%%%%%%%%%%%%%%%%%%%%%%%%

\section{Introduction}
\label{sec:introduction}

It is expected that dynamics of wormholes are one of the most essential ingredients in understanding nature of fully quantum gravity. 
Dynamical wormholes are particular to quantum gravity and, therefore, they would sometimes provide clues to issues that quantum gravity addresses 
even at a semiclassical level. For example, the replica wormhole plays an important role in analyzing the information loss problem in the entropy 
of the Hawking radiation \cite{Penington:2019kki}. 
Another issue associated with dynamical wormholes comes from the proposal 
by Coleman on the tiny cosmological constant \cite{Coleman:1988tj}. 
It is quite interesting because of not only the role played by dynamical wormholes, 
but its relation to averaging nature of quantum gravity \cite{Saad:2019lba, Maloney:2020nni}. 
Hence it is desirable to study nature of dynamical wormholes in a well-defined 
set up of quantum gravity beyond semiclassical analysis (see, e.g., Refs.~\cite{Ambjorn:2021wdm, Hamada:2022dan}).  
In this article we examine $2$D quantum gravity in such a physical model. 

The non-perturbative physics of $2$D quantum gravity has been extensively studied with the help of matrix models and combinatorics that provide quite powerful tools for analytical calculations (see, e.g., Refs.~\cite{DiFrancesco:1993cyw,Ambjorn:1997di, Ambjorn:2022mxb} for reviews). 
Matrix models can be interpreted as the lattice models of random surface, known as dynamical triangulations \cite{Ambjorn:1985az,Ambjorn:1985dn,David:1984tx,Billoire:1985ur,Kazakov:1985ea,Boulatov:1986jd}, 
and serve as well-defined regularizations of $2$D quantum gravity coupled to conformal matter with the central charge $c$ less than or equal to $1$.

The simple example is given by the $N\times N$ Hermitian one-matrix model with the quartic single-trace interaction, which was first solved in the large-$N$ limit in the seminal paper \cite{Brezin:1977sv}. 
This matrix model corresponds to the model of random surface discretized by quadrangles. 
Tuning the coupling for the quartic interaction $g$ to its critical value $g_c$, one can take the continuum limit through which the cutoff is removed. 
This critical behavior is quantified by the string susceptibility exponent $\gamma = -1/2$.  
The resulting continuum theory is the pure $2$D quantum gravity, known as the Liouville field theory coupled to conformal matter with $c=0$.

An interesting modification of the matrix model above is to introduce the double-trace interaction \cite{Das:1989fq}. 
In this article, we call the matrix model having the double-trace interaction the double-trace matrix model. 
From the point of view of the random surface, the double-trance interaction can control the weight for the surfaces touched through tiny necks which are microscopic wormholes. 
One can take the continuum limit approaching the critical line $g_c(g_D)$ where the $g_D$ is the coupling for the double-trace interaction. 
When $g_D$ is smaller than the special value $g_D^*$, the continuum theory is the pure $2$D quantum gravity characterized by $\gamma = -1/2$. 
At $g_D=g_D^*$, the effect of wormholes becomes important enough to change the critical behavior, and the string susceptibility turns to the positive value $\gamma = + 1/3.$\footnote{
The positive string susceptibility $\gamma = +1/3$ has also been obtained by the modified multicritical one-matrix model with the single-trace interaction \cite{Ambjorn:2016lkl, Ambjorn:2017ibv}. 
}  
When $g_D > g_D^*$, one enters the branched-polymer phase with $\gamma = +1/2$ where the surfaces become ``many-fingered'' polymer-like chains.     

As pointed out in Ref.~\cite{Klebanov:1994pv}, the continuum theory with the string susceptibility $\gamma = +1/3$ 
can be interpreted as the Liouville field theory that is coupled to conformal matter with $c=0$ and defined by the unconventional branch of the gravitational dressing in the Liouville potential.  
This type of theory is not smoothly connected to the semiclassical regime $c\to - \infty$.   
It is generically suggested that the conventional Liouville field theory with the negative string susceptibility $\gamma$ 
can be related to the unconventional Liouville field theory\footnote{
The unconventional Liouville field theory means the Liouville field theory defined by the unconventional branch of the gravitational dressing in the Liouville potential. 
} with the positive string susceptibility $\gamma/(\gamma - 1)$ 
by the fine tuning of the coupling for the double-trace interaction in the matrix model \cite{Klebanov:1994pv, Klebanov:1994kv}. 
The same formula for the string susceptibility was found in Ref.~\cite{Durhuus:1994tu} based on the combinatorial argument.

In this article, we wish to study the double-trace matrix model in the large-$N$ limit, and investigate the continuum limit in which the wormhole effects become substantial. 
In particular, we calculate the continuum limit of the resolvent which is the marked disk amplitude. 
The resulting continuum disk amplitude is exactly the same as that of the pure $2$D quantum gravity although $\gamma = + 1/3$.  
We also introduce the renormalized coupling for the double-trace interaction, and show that the newly introduced coupling can shift the renormalized bulk cosmological constant effectively, 
which is essentially caused by the dominance of microscopic wormholes. 
As a result, the effective bulk cosmological constant can be zero even if the original bulk cosmological constant is positive.

This article is organized as follows. 
In Sec.~\ref{eq:doubletracemm}, we give a brief introduction to the double-trace matrix model and explain its relation to the lattice model of random surface. 
In Sec.~\ref{sec:largenlimit}, the physics at large $N$ is explained, and we show the three possible critical behaviors. 
Additionally we examine the free energy near the critical point where the wormholes become important. 
In Sec.~\ref{sec:continuumlimit}, we calculate the continuum limit of the disk amplitude at the critical point characterized by $\gamma = +1/3$, 
and then introduce the renormalized coupling for the double-trace interaction. 
In Sec.~\ref{sec:NE}, we derive nonperturbative effect in the continuum limit.  
Sec.~\ref{sec:discussion} is devoted to the discussion.

\section{Overview of the double-trace matrix model}
\label{eq:doubletracemm}
We study the Hermitian one-matrix model with double-trace term, given by the matrix integral \cite{Das:1989fq}:
\[
Z_N(g, g_D) 
=
\frac{1}{\Omega_N}
\int D\phi\
e^{ -N \left[ \text{tr}\, V\left(\phi\right) - \frac{g_D}{2N} \left( \text{tr}\, W\left(\phi\right)\right)^2 \right]}\ , \ \ \ \text{with}\ \ \
\Omega_N = \frac{(2\pi)^{\frac{N(N-1)}{2}}}{G_2(N+2)}\ , 
\label{eq:dtmm}
\]
where $\phi$ is an $N\times N$ Hermitian matrix, and $D\phi$ the Haar measure on a Hermitian matrix: 
\[
D\phi = 2^{\frac{N(N-1)}{2}} 
\prod^{N}_{i=1} \text{d} \phi_{ii} \prod_{1\le j < k \le N} \text{d} \text{Re}\left( \phi_{jk} \right) \text{d} \text{Im}\left( \phi_{jk} \right)\ ,
\label{eq:measure}
\]  
and $G_2 (z)$ is the Barnes function defined by
\[
G_2 (z+1) = \Gamma (z) G_2 (z)\ , \ \ \ G_2(1) = 1\ . 
\label{eq:barnes}
\]
Here $V(\phi)$ and $W(\phi)$ are the polynomial and the monomial, respectively:
\[
V(\phi) = \frac{1}{2} \phi^2 - g \sum^{k}_{m=1} \frac{t_m}{m} \phi^m\ ; \ \ \ W(\phi) =  \phi^n\ , \ \ \ \text{with}\ \ \ n \in \mathbb{N}_+\ , 
\label{eq:vandw}
\]
where $t_m$'s fulfill $t_1, t_2, \cdots, t_{k-1} \ge 0$ and $t_k > 0$. 
Based on the notation above, the Gaussian matrix integral yields  
\[
Z_N(0,0) = \frac{1}{\Omega_N} \left( \frac{ 2\pi }{ N } \right)^{\frac{N^2}{2}}\ . 
\label{eq:gaussianZ}
\]
In the following, we briefly review the matrix model with double-trace term and its relation to $2$D quantum gravity.  

\subsection{Pure gravity}
\label{sec:puregravity}

Let us first assume that $g_D = 0$. 
The perturbative expansion of the integral (\ref{eq:dtmm}) w.r.t. $g$ can be interpreted as a model of random surface, 
i.e. the dual graph of each Feynman graph is a complex made up of the $m$-sided polygons ($m=1,2, \cdots , k$), 
and in this sense the sum over Feynman graphs defines the lattice model of random surface.  
For simplicity, we set $t_4=1$ and other $t_m$'s are zero, and the perturbative expansion then yields  
\[
\log  \left( \frac{Z_N(g,0)}{Z_N(0,0)} \right)
\cong 
\sum_G \frac{N^{\chi(G)}}{|\text{Aut} (G)|} g^{n(G)}\ , 
\label{eq:perturbative}
\] 
where $G$ denotes a quadrangulated, closed, connected, and oriented surface, 
$|\text{Aut} (G)|$ the order of automorphism group of $G$, $\chi (G)$ the Euler characteristic of $G$, 
and $n(G)$ the number of quadrangles in $G$. It is the logarithm that selects the connected surfaces. 
The symbol $\cong$ indicates equality up to perturbative expansion. 
Hereafter, we use the symbol $\cong$ with this meaning unless stated otherwise.  

The model of random surface (\ref{eq:perturbative}) serves as a regularization of $2$D quantum gravity, 
meaning that each quadrangle is a square with the lattice spacing $\varepsilon$, which is the UV cutoff. 
In the context of $2$D quantum gravity, $g$ and $N$ are interpreted as the (dimensionless) cosmological constant $\lambda$ and the gravitational constant $\kappa$ through the relation:
\[
g = e^{-\lambda}\ , \ \ \ N = e^{1/\kappa}\ . 
\label{eq:lambdakappa}
\]   

$2$D quantum gravity is known to be asymptotically free due to e.g. the $\epsilon$-expansion \cite{Kawai:1993mb}. 
Therefore, to remove the cutoff, one has to tune the bare gravitational constant $\kappa$ to $0$, or equivalently the matrix size $N$ to infinity.  
If taking the large $N$ limit first, the surfaces with the spherical topology survive in the sum, and Eq.~(\ref{eq:perturbative}) then becomes
\[
\lim_{N \to \infty} \frac{1}{N^2} \log  \left( \frac{Z_N(g,0)}{Z_N(0,0)} \right) 
\cong 
\sum_{G_0} \frac{1}{|\text{Aut} (G_0)|} e^{- \lambda n(G_0)}\ , 
\label{eq:perturbative2}
\] 
where $G_0$ denotes a quadrangulated, closed, connected, and oriented surface homeomorphic to the sphere. 
If replacing the sum over quadrangulations by the one over the number of quadrangles, 
\[
\sum_{G_0} \frac{1}{|\text{Aut} (G_0)|} e^{- \lambda n(G_0)}
= \sum_n e^{-\lambda n} \mathcal{N}(n)
\ , 
\label{eq:perturbative3}
\] 
where $\mathcal{N}(n)$ roughly counts the number of quadrangulations for a given $n$. 
When the number of quadrangles is large, $\mathcal{N}(n)$ behaves as
\[
\mathcal{N}(n) \propto e^{\lambda_c n} n^{\gamma - 3} \left( 1 + \mathcal{O} \left( 1/n \right) \right)\ ,\ \ \ \text{with}\ \ \ \gamma = - \frac{1}{2}\ ,  
\label{eq:asymptotic}
\] 
where $\lambda_c$ is a certain constant and $\gamma$ is called the string susceptibility.   
The important fact is that $\gamma$ is universal, while $\lambda_c$ is not. 
The sum (\ref{eq:perturbative3}) is therefore exponentially bounded, and hits the radius of convergence at $\lambda = \lambda_c$. 
From Eq.~(\ref{eq:asymptotic}), tuning $\lambda$ to $\lambda_c$ from above, infinitely many quadrangles become important in the sum (\ref{eq:perturbative3}), 
and essentially the average number of quadrangles diverges. 
Therefore, if we simultaneously tune the lattice spacing $\varepsilon$ to $0$ in a correlated manner, we may obtain the continuous surface. 
This is the essence of the continuum limit that removes the cutoff. 
Specifically, we tune $\lambda \searrow \lambda_c$ and $\varepsilon \searrow 0$ with the renormalized cosmological constant $\Lambda$ kept fixed: 
\[
\Lambda =
\frac{\lambda - \lambda_c}{\varepsilon^2} \ , 
\]
where the bare cosmological constant $\lambda / \varepsilon^2$ gets the additive renormalization. 
Through this dimensional transmutation, one can transmute the dimension of lattice spacing to the dimension of the renormalized cosmological constant, and set the scale at IR. 
In the continuum limit, one obtains    
\[
e^{-\lambda n} \mathcal{N}(n) \sim e^{- \Lambda A} A^{\gamma - 3}\ , 
\label{eq:liouvilleresult}
\] 
where the physical area $A$ is defined to be finite as $A:= \varepsilon^2 n$. 
The symbol $\sim$ indicates that unimportant numerical factors are ignored, and an important universal part is selected. 
Hereafter, we use the symbol $\sim$ with this meaning unless stated otherwise.    
In fact, the result (\ref{eq:liouvilleresult}) can be recovered by the path-integral of $2$D quantum gravity, the Liouville quantum gravity, with the fixed area $A$.  

We stress here that the critical behavior of the sum (\ref{eq:perturbative3}) can be quantified by the string susceptibility $\gamma$:
\[
\sum_{G_0} \frac{1}{|\text{Aut} (G_0)|} e^{- \lambda n(G_0)} \sim (g_c - g)^{2 - \gamma}\ , \ \ \ \text{with}\ \ \ g_c := e^{-\lambda_c} = \frac{1}{12}\ . 
\label{eq:gamma}
\] 
Essentially one can observe this critical behavior through the free energy as well:
\[
F(g,g_D=0) = - \lim_{N \to \infty} \frac{1}{N^2} \log Z_N (g,g_D=0) \sim (g_c - g)^{2-\gamma}\ . 
\label{eq:free_energy_sing_pure}
\]
The string susceptibility $\gamma$ is universal in a sense that even if one studies the model of random surface discretized by generic polygons, the same value of the exponent can be obtained in the continuum limit.

\subsection{Introduction of wormholes}
\label{sec:introductionofwormholes}

Turning on the coupling $g_D$, one can introduce tiny necks that connect distinct points on surfaces. 
The tiny necks are microscopic wormholes. 
Introducing the double-trace coupling explicitly, certain ``disconnected'' surfaces, i.e. surfaces touched through wormholes, 
will survive even after taking the logarithm. 

The free energy in the large-$N$ limit, 
\[
F(g,g_D) = - \lim_{N \to \infty} \frac{1}{N^2} \log  Z_N (g, g_D)  \ , 
\label{eq:free_energy}
\]
is known to become singular if tuning the coupling $g$ for a given value of $g_D$ to its critical value $g_c(g_D)$: 
\[
F(g,g_D) \sim ( g_c (g_D) - g )^{2 - \gamma (g_D)}\ , 
\label{eq:gamma2}
\]
where the critical exponent $\gamma$ is the string susceptibility that depends on the double-trace coupling.  
It is shown in Ref.~\cite{Das:1989fq} that there exists a critical point of $g_D=g_D^*$ 
by which we have three phases. 
For $- \infty <  g_D <  g_D^*$, one obtains the pure-gravity phase characterized by $\gamma = -1/2$. 
This means that the effect of wormholes is not strong enough to change the critical behavior obtained by the model without the double-trace term. 
For $g_D > g_D^*$, one enters the branched-polymer phase where surfaces degenerate into the tree-like structure, and this phase is characterized by $\gamma = +1/2$. 
In between these phases, i.e., at $g_D=g_D^*$, the string susceptibility becomes $\gamma = +1/3$. 
At this special point on the critical curve, the effect of wormholes becomes substantial to change the critical exponent. In this article, we call this phase the wormhole-dominant phase.

Another interesting aspect of the matrix model with double-trace term was pointed out \cite{Klebanov:1994kv}. 
Performing the Hubbard-Stratonovich transformation, the matrix integral (\ref{eq:dtmm}) becomes    
\[
Z_N(g, g_D) 
=  
\frac{1}{\Omega_N}
\int D\phi 
\left\langle 
e^{-N \text{tr} \left( V(\phi)+\nu W(\phi) \right)}
\right\rangle_{G} 
\ ,
\label{eq:KH}
\]
where the average is defined by  
\[
\langle f(\nu) \rangle_{G} = 
\int^{\infty}_{-\infty}\text{d} \nu\ \rho_G (\nu) f(\nu)\ . 
\label{eq:average}
\]
Here $\rho_G (\nu)$ is the Gaussian probability density: 
\[
\rho_G (\nu) = \frac{1}{\sqrt{2\pi} \sigma}\ e^{- \frac{\nu^2}{2\sigma^2}}\ , \ \ \ \text{with}\ \ \ \sigma = \frac{\sqrt{g_D}}{N}\ .
\label{eq:gaussian}
\]
Therefore, one can recover the original matrix integral (\ref{eq:dtmm}), 
if one starts with the one-matrix model only with the single-trace term  
by performing the Gaussian integration over its coupling. 

Let us formally interchange the order of integrations in Eq.~(\ref{eq:KH}) when $N$ is large but finite. 
If we assume that $V(\phi) = \frac{1}{2}\phi^2 - \frac{g}{4}\phi^4$ and $W(\phi) =  \phi^4$,
Eq.~(\ref{eq:KH}) may yield
\[
e^{- \mathcal{F} (g,g_D)} =  
\int^{\infty}_{- \infty}\text{d} \nu\ \rho_G(\nu)\ e^{-  \mathcal{F}(g+4\nu, 0)}\ , \ \ \ \text{with}\ \ \ \mathcal{F} (g,g_D) :=  - \log Z_N (g,g_D)\ . 
\label{eq:KH2}
\]  
This implies the average over the coupling constant, which is reminiscent of the Coleman mechanism \cite{Coleman:1988tj}.
However, the integral (\ref{eq:KH2}) seems problematic even in the large-$N$ limit:  
The free energy in the large-$N$ limit, $F(g,g_D)$, has a finite radius of convergence with respect to $g$, 
and therefore the integration over $\nu$ will diverge for large $|\nu|$ \cite{Hamada:2022dan}.

Assuming that we can somehow regularize the integral (\ref{eq:KH2}), let us formally proceed calculations. 
Changing the variable $\nu$ to $t$, 
\[
g + 4 \nu = \frac{1}{12} \left( 1 - \varepsilon^2 t \right)\ ,
\label{eq:nutot}
\] 
the integral (\ref{eq:KH2}) accordingly becomes
\[
\int^{\infty}_{-\infty}\text{d} \nu\ \rho_G(\nu)\ e^{- \mathcal{F}(g+4\nu, 0)}
\propto 
\int^{ \infty }_{ -\infty } \text{d}t\ e^{ f(t)}\ ,
\label{eq:regularization2}
\] 
where 
\[
f(t) \cong  N^2 \left(f_1\, \varepsilon^2 t 
+ f_2\,  (\varepsilon^2 t)^2 
+ f_{5/2}\, (\varepsilon^2t)^{5/2} + \mathcal{O}((\varepsilon^2t)^{3})
\right) 
+ \mathcal{O}(N^0)
\ .
\label{eq:ft}
\]
Here the coefficients are
\[
f_1 = \frac{1-12g - 2^6 \cdot 3 g_D}{2^8 \cdot 3^2}\ , \ \ \ 
f_2 = - \frac{  1 - 2^6 \cdot 3^2 g_D }{2^9 \cdot 3^2 g_D}\ , \ \ \ 
f_{5/2} = - \frac{4}{15}\ . 
\label{eq:fs}
\]
The three phases in the double-trace matrix model can be characterized by these coefficients \cite{Klebanov:1994kv}: 
$(i)$ $f_1 = 0$ and $f_2 < 0$ yield the pure-gravity phase, $(ii)$ $f_1 = 0$ and $f_2 > 0$ the branched-polymer phase, 
and $(iii)$ $f_1 = f_2 = 0$ the wormhole-dominant phase.  

In order to explore the possibility to discuss the Coleman mechanism, 
let us parametrize the couplings as follows: 
\[
g = g_* (1-6\varepsilon^3\ \overline{t}) = \frac{1}{18}(1-6\varepsilon^3\ \overline{t})\ , \ \ \ g_D = g^*_D = \frac{1}{2^6 \cdot 3^2}\ , 
\label{eq:ggdparametrization}
\]
where $g_* (:= g_c(g^*_D))$ is the critical coupling for the wormhole-dominant phase.
Accordingly, we obtain
\[
f(t) \cong  N^2 \left( \left( t \overline{t} - \frac{4}{15} t^{5/2} \right) \varepsilon^5 + \mathcal{O}(\varepsilon^6) \right) + \mathcal{O}(N^0)\ , 
\label{eq:ft2} 
\]
From the observation above, one may guess that one can take the double-scaling limit with $N^2 \varepsilon^{5}$ kept fixed. 
In fact, if taking the limit $N\to \infty$ with $N \equiv \varepsilon^{-5/2}$ and removing the regularization properly, 
the integral (\ref{eq:KH2}) essentially becomes the two-sided Laplace transform \cite{Klebanov:1994kv}\footnote{We choose the unit such that $g_s = N \varepsilon^{5/2} = 1$.}:
\[
e^{ - \overline{\mathcal{F}}(\overline{t}) }
= \int^{\infty}_{-\infty} \text{d}t\ e^{  t \overline{t} - \mathcal{F} (t) } \ , 
\label{eq:Laplace}
\]
where $\overline{\mathcal{F}}( \overline{t} )$ and $\mathcal{F} (t)$ are respectively the free energies for the pure-gravity and wormhole-dominant phases in the double-scaling limit. 
As mentioned in Ref.~\cite{Klebanov:1994kv}, it is unclear if the integral (\ref{eq:Laplace}) converges and if it gives the truly non-perturbative formulation\footnote{The non-perturbative abmiguity was pointed out in Ref.~\cite{Diego:1995my}}.   
However, the point is that Eq.~(\ref{eq:Laplace}) works at all orders of perturbation theory \cite{Klebanov:1994kv}. 
Therefore, there is a possibility to discuss the Coleman mechanism using Eq.~(\ref{eq:Laplace}), which is quite interesting.

In Ref.~\cite{Das:1989fq} the critical point of $g_D$ can be obtained by requiring divergence of the susceptibility in the case of $V(\phi)=\frac12\phi^2-\frac{g}{4}\phi^4$, 
$W(\phi)=\phi^2$. 
On the other hand, in Ref.~\cite{Klebanov:1994kv} the critical point is chosen to satisfy $f_1=f_2=0$ in the case of $V(\phi)=\frac12\phi^2-\frac{g}{4}\phi^4$, $W(\phi)=\phi^4$ as we have seen above.  
It is likely that the equivalence of these two approaches is ensured at least perturbatively by the fact that both of them gives Eq.~(\ref{eq:Laplace}) in the double-scaling limit, 
but a prior it is not clear, in particular, that the former leads to Eq.~(\ref{eq:Laplace}). 
Thus in the next section we confirm that they actually define the same double scaling limit even in the case of different $W(\phi)$'s from the original ones discussed in Refs.~\cite{Das:1989fq,Klebanov:1994kv}. 
Of course this check is far from a proof of their equivalence, but at least it should be necessary to compare their prescriptions in other examples to examine their relationship.

\section{Large-$N$ limit}
\label{sec:largenlimit}
Let us first review the large-$N$ physics of the double-trace matrix model based on the pioneering paper \cite{Das:1989fq}.

\subsection{Free energy}
\label{sec:freeenergy}
We wish to calculate the free energy in the large-$N$ limit to extract the critical behaviors. 

The matrix integral (\ref{eq:dtmm}) is invariant under the U($N$) transformation, i.e., $\phi \to U \phi U^{\dagger}$ where $U$ is an $N\times N$ unitary matrix. 
Through the use of transformation under the U($N$) group, one can diagonalize the matrix $\phi$: 
\[
\phi \to U \phi U^{\dagger} = \text{diag} (\lambda_1, \lambda_2, \cdots , \lambda_N)\ . 
\label{eq:diagonalization}
\]
Accordingly, the integral measure (\ref{eq:measure}) becomes
\[
D\phi = \Omega_N \prod^{N}_{i=1} \text{d}\lambda_i \prod^N_{j<k} | \lambda_j - \lambda_k |^2\ . 
\label{eq:measure2}
\]
The matrix integral (\ref{eq:dtmm}) then can be recast into the multiple integration over eigenvalues:
\[
Z_N(g,g_D) 
=
\int^{\infty}_{-\infty} \cdots \int^{\infty}_{-\infty}  \prod^{N}_{i=1}d\lambda_i\ e^{-N^2 V_{\text{eff}}(\lambda)}\ , 
\label{eq:dtmm2}
\] 
where 
\[
V_{\text{eff}}(\lambda) 
= \frac{1}{N}\sum^{N}_{i=1} V(\lambda_i)
-\frac{g_D}{2}\left(\frac{1}{N}\sum^{N}_{i=1} W(\lambda_i)\right)^2 
- \frac{1}{N^2}\sum^N_{i=1}\sum^N_{j\ne i} \log |\lambda_i - \lambda_j| \ .
\label{eq:veff}
\]
Introducing 
\[
\rho(\lambda)=\frac1N\text{tr}\,\delta(\lambda-\phi)=\frac1N\sum^{N}_{i=1}\delta(\lambda-\lambda_i)\ ,\] 
this becomes 
\[
V_{\text{eff}}
= \int d\lambda\,\rho(\lambda\,) V(\lambda)
-\frac{g_D}{2}\left(\int d\lambda\,\rho(\lambda)W(\lambda)\right)^2 
- \int d\lambda \dashint d\mu\,\rho(\lambda)\rho(\mu)\log |\lambda-\mu| \ .
\label{eq:Veffbyrho} 
\]
In the large-$N$ limit, we solve the saddle point equation for $\rho(\lambda)$ 
\[
&V(\lambda)-g_DW_0W(\lambda)-2\dashint d\mu\,\rho(\mu)\log|\lambda-\mu|+C=0\ ,
\label{eq:eqforrho} \\
&W_0:=\int d\mu\,\rho(\mu)W(\mu)\ ,
\label{eq:eqforW0}
\]
where we have introduced the multiplier $C$ to impose $\int d\lambda\,\rho(\lambda)=1$. 
Hence 
\[
V'(\lambda)-g_DW_0W'(\lambda)
-2\dashint d\mu\,\frac{\rho(\mu)}{\lambda-\mu}\,=0\ .
\label{eq:spe}
\]
Thus in order to get $\rho(\mu)$, we solve this equation with the self-consistency condition \eqref{eq:eqforW0}. 
Conversely, suppose we find a solution to this equation $\rho_0(\lambda)$ and $V(0)=W(0)=0$, integrating Eq.~\eqref{eq:spe} gives 
\[
V(\lambda)-g_DW_0W(\lambda)
-2\dashint d\mu\,\rho_0(\mu)\left(\log|\lambda-\mu|-\log|\mu|\right)\,=0\ .
\]
Plugging this back into Eq.~\eqref{eq:Veffbyrho}, we get the free energy in the large-$N$ limit:   
\[
F_0(g,g_D)
=\int d\lambda\,\rho_0(\lambda)\left(\frac12V(\lambda)-\log|\lambda|\right)\ , 
\label{eq:Fbyrho}
\] 
which is actually the ``same'' equation as the one obtained in Ref.~\cite{Brezin:1977sv}. 
Thus we find that the expression of the free energy in terms of the solution to the saddle point equation takes exactly the ``same'' form 
as in the model without the double-trace term, but we notice that the eigenvalue distribution $\rho_0(\lambda)$ does change due to its presence as shown in  
Eq.~\eqref{eq:spe}.

Introducing the resolvent, 
\[
R_0(z) = \vev{\frac{1}{N} \text{tr} \left( \frac{1}{z - \phi} \right)}_0\ , 
\label{eq:resolvent}
\]
where $\vev{\cdot}_0$ denotes the expectation value in the large-$N$ limit 
in the model \eqref{eq:dtmm}, 
the saddle-point equation (\ref{eq:spe}) can be recast as the loop equation:
\[
R_0(z)^2 - \widetilde V'(z) R_0(z) + Q_0(z) = 0\ ,
\label{eq:loopeq}
\]
where 
\[
\widetilde V(z):=V(z)-g_DW_0W(z)\ , \ \ \ 
\text{with}
\ \ \ 
W_0=\vev{\frac{1}{N}\text{tr}\,W(\phi)}_0\ , 
\label{eq:W0}
\]
and $Q_0(z)$ is a polynomial of $z$ with degree $\text{max}\{\text{deg }V, \text{deg }W\}-2$: 
\[
Q_0(z) = \vev{\frac{1}{N}\frac{\widetilde V'(z) - \widetilde V'(\phi)}{z-\phi}}_0 \ . 
\label{eq:q}
\]
In App.~\ref{app:SD} we give another derivation of Eq.~\eqref{eq:loopeq}
 based on the Schwinger-Dyson equation. 
In solving Eq.~\eqref{eq:loopeq}, 
we assume that the resolvent has a single cut. 
Since it is quadratic, there generically exist two solutions and we pick up the one 
consistent with the asymptotic behavior of the resolvent, i.e., $R_0(z)\cong 1/z + \mathcal{O}(z^0)$ for $|z|\gg1$. 
The solution to the loop equation (\ref{eq:loopeq}) then yields
\[
R_0(z) = \frac{1}{2} 
\left(
\widetilde V'(z)
+f(z) \sqrt{(z-a_1)(z-a_2)}
\right)\ , 
\label{eq:solutionofw}
\]
where $a_1$ and $a_2$ ($a_1<a_2$) are endpoints of the cut, and $f(z), a_1, a_2$ are chosen 
in such a way that 
\[
&f(z)^2(z-a_1)(z-a_2)=\widetilde V'(z)^2-4Q_0(z)\ , \nonumber \\
&R_0(z)\cong 1/z + \mathcal{O}(z^0) \ , \ \ \ \text{for}\ \ \ |z|\gg1\ . 
\label{eq:acond}
\]
Note here that $W_0$ should satisfy the condition (\ref{eq:W0}).

The resolvent,  
\[
R_0(z) = \int^{a_2}_{a_1} \text{d}\mu\ \frac{\rho_0 (\mu)}{z - \mu}\ ,
\ \ \ \text{with} \ \ \ \rho_0(\mu)=\vev{\rho(\mu)}_0\ ,  
\label{eq:resolvent2}
\]
is analytic except for the range $z \in \left(a_1, a_2 \right)$, and  
the eigenvalue density can be read off through the non-analytic part of the resolvent, i.e., for $z\in[a_1,a_2]$, 
\[
\rho_0 (\lambda) 
= \frac{1}{2\pi i} 
\left(
R_0(\lambda + i 0) - R_0(\lambda - i 0)
\right)
= \frac{1}{2\pi} 
f(\lambda)
\sqrt{(\lambda-a_1)(a_2-\lambda)}\ ,
\label{eq:rho2}
\]
where we have used Dirac's delta function defined by  
\[
\delta (\lambda) &= \frac{-1}{2\pi i} \left( \frac{1}{\lambda + i  0} - \frac{1}{\lambda -i0} \right)\ . 
\label{eq:delta}
\]
Inserting Eq.~(\ref{eq:rho2}) into Eq.~(\ref{eq:eqforW0}), 
the self-consistency equation 
\[
W_0=\int d\mu\,\rho_0(\mu)W(\mu)\ ,
\label{eq:consistencycond}
\]
provides the relation between $W_0$ and $a_1$, $a_2$. 

As illustrative examples, let us fix $V(z)=\frac12z^2-\frac g4z^4$ and 
consider the case 
$W(z)=z^2$ or $W(z)=z^4$, which has been studied in Ref.~\cite{Das:1989fq} and Ref.~\cite{Klebanov:1994kv}, respectively. 
In these cases, we can set $-a_1=a_2=a>0$. 

For $W(z)=z^2$, the resolvent \eqref{eq:solutionofw} takes the form 
\[
R_0(z) = \frac{1}{2} 
\left(
\left(1- 2g_D W_0\right)z 
-gz^3 
+ \left( g z^2 + \frac{1}{2}a^2 g - (1- 2 g_D W_0) \right) 
\sqrt{z^2 - a^2}
\right)\ , 
\label{eq:R0forW2}
\]
where Eq.~\eqref{eq:acond} makes 
$a$ subject to the quartic equation:  
\[
3g a^4 - 4 \left( 1- 2 g_D W_0 \right) a^2 + 16 = 0
\label{eq:a4}
\ . 
\]
Eq.~\eqref{eq:rho2} leads to 
\[
\rho_0 (\lambda) 
= \frac{1}{2\pi} 
\left(
 -g\lambda^2 - \frac{g}{2}a^2 + 1 - 2g_D W_0
\right)
\sqrt{a^2-\lambda^2}\ . 
\label{eq:rhoforW2}
\]
From Eq.~\eqref{eq:consistencycond}, we also obtain the self-consistent equation for $W_0$:  
\[
W_0 =  \int^a_{-a} \text{d} \lambda\ \rho (\lambda) \lambda^2 = \frac{a^4}{16}\left( 1-ga^2 - 2g_D W_0 \right)\ .
\label{eq:w22}
\]
If solving for $W_0$ in Eq.~(\ref{eq:w22}), we get 
\[
W_0 = \frac{a^4-ga^6}{16+ 2g_D a^4}\ . 
\label{eq:w2consistent}
\]
Plugging Eq.~(\ref{eq:w2consistent}) into Eq.~(\ref{eq:a4}), one can eliminate $W_0$, and find the equation for $a$ itself: 
\[
g g_D  a^8  - 8 (3g + 2g_D )a^4 + 32 a^2- 128 = 0\ . 
\label{eq:eighthorderequation}
\]
Note that in the limit $g_D\rightarrow 0$ this equation correctly reproduces 
the result in \cite{Brezin:1977sv}\footnote
{$a$ in the present paper is half of one in \cite{Brezin:1977sv}.}:
\[3ga^4-4a^2+16=0\ .
\label{eq:aeq}
\] 

Finally, from Eq.~\eqref{eq:Fbyrho} the free energy in the large-$N$ limit is given as 
\[
F_0(g,g_D) 
&=\int^a_0 \text{d}\lambda\ \rho (\lambda)
\left( 
\frac{1}{2} \lambda^2 - \frac{g}{4} \lambda^4 - 2 \log |\lambda|
\right)  \notag \\
&=\frac{a^2}{2048}
\biggl[
9 g^2 a^6 - 8g (5-2g_D W_0) a^4 \notag \\
&\ \ \ + 32 \biggl( 1 -2g_D W_0 -3g +12g \log\left(\frac a2\right) \biggl)a^2 \notag \\
&\ \ \ + 256 (1-2g_D W_0) \left( 1 - 2 \log\left(\frac a2\right) \right)
\biggl] 
\label{eq:F0inW0}\\
&=
 \frac{ g a^4 \left( 3g a^4 - 8a^2 + 64 \right)   }{2048} 
 + \frac{1}{16} a^2 - \log\left(\frac a2\right) + \frac{1}{2}  
\ ,
\label{eq:free_energy2}
\]
where $a$ is subject to Eq.~(\ref{eq:eighthorderequation}) and hence is a function of $g$ and $g_D$.
In the last line, we have used Eq.~(\ref{eq:a4}) to eliminate $W_0$. 
We can confirm that taking the limit $g_D\rightarrow 0$ in Eq.~\eqref{eq:F0inW0} 
and using Eq.~\eqref{eq:eighthorderequation} in this limit, 
the free energy of the single-trace one-matrix model given in Ref.~\cite{Brezin:1977sv}, 
\[
F_0 =\frac{30g-1}{288g}a^2+\frac{1}{72g}+\frac38-\log\left(\frac a2\right)\ , 
\label{eq:BIPZresult}
\]
is recovered. In this sense, the two limits, $N\rightarrow\infty$ and $g_D\rightarrow 0$, commute. 

The free energy for $W(z)=z^4$ is obtained in the same way. 
The resolvent and the eigenvalue distribution read
\[
&R_0(z)= \frac{1}{2} 
\left(
z-\left(g + 4g_D W_0\right)z^3 
+ \left(g + 4 g_D W_0\right)(z^2-b^2) 
\sqrt{z^2 - a^2}
\right)\ , \nonumber \\
&\rho_0(\lambda)=\frac{1}{2\pi} 
\left(g + 4 g_D W_0\right)(b^2-\lambda^2) 
\sqrt{a^2 - \lambda^2}\ ,
\label{eq:R0forW4}
\]
where 
\[
b^2=\frac{1}{g+4g_DW_0}-\frac12a^2\ , 
\label{eq:bbya}
\]
and $a$ satisfies 
\[
3(g+4g_DW_0)a^4-4a^2+16=0\ .
\]
The self-consistency condition tells us that 
\[
W_0 = \frac{a^6(8-9ga^2)}{256+ 36g_D a^8}\ , 
\]
which yields a closed equation for $a$ as 
\[
3 g_D  a^{10}  -36 g_Da^8-48g a^4 + 64 a^2- 256 = 0\ . 
\label{eq:tenthorderequation}
\]
It is again reduced to Eq.~\eqref{eq:aeq} in the limit $g_D\rightarrow 0$. 
We obtain the free energy as 
\[
F_0&=\frac{a^2}{2048}\left(g+4g_DW_0\right)
\biggl[
5 g a^6 - 8g a^4 b^2-16a^4+32a^2b^2+32a^2+256b^2
+ 128 \left( a^2-4b^2 \right)\log\left(\frac a2\right) \notag 
\biggl] \notag \\
&=\frac{1}{32(9g_Da^8+64)}
\left(9\left(g^2+40g_D\right) a^8 -32 ga^6+64 \left( 6g + \frac13  \right)a^4-\frac{1024}{3}a^2+2560\right) - \log\left(\frac a2\right) \ ,
\] 
which becomes Eq.~\eqref{eq:BIPZresult} in the limit $g_D\rightarrow 0$ due to Eq.~\eqref{eq:aeq}.

\subsection{Critical phenomena}
\label{sec:ctiricalphenomena}

In order to find the critical phenomena for $W(z)=z^2$, we solve for $g$ in Eq.~(\ref{eq:eighthorderequation}): 
\[
g = \frac{ 16 ( 8 + \zeta (g_D \zeta - 2) ) }{ \zeta^2 (g_D \zeta^2  - 24 ) }\ , 
\label{eq:gofzeta}
\]
where $\zeta := a^2$. Using Eq.~(\ref{eq:free_energy2}) and Eq.~(\ref{eq:gofzeta}), 
one can compute the susceptibility:
\[
\chi 
= \frac{\partial^2}{\partial g^2} F \biggl{|}_{g_D}
= \frac{\partial^2 F}{\partial \zeta^2} \left( \frac{\text{d} g}{\text{d} \zeta}\right)^{-2} 
- \frac{\partial F}{\partial \zeta}  \frac{\text{d}^2 g}{\text{d} \zeta^2} \left(\frac{\text{d} g}{\text{d} \zeta}\right)^{-3}
= -\frac{\zeta^4}{1024} \left( 
\frac{ g_D \zeta^2 - 72  }{ g_D \zeta^2 - 8 } 
\right)\ .  
\label{eq:chi}
\]   
The singularities of the susceptibility comes from the zeros of the denominator, 
and the roots of $\text{d}g/\text{d}\zeta = 0$ \cite{Das:1989fq}. 
The latter means that one cannot obtain $\zeta (g)$ by the inversion.   
Here
\[
\frac{ \text{d} g}{ \text{d} \zeta} 
= - \frac{ 32 ( g_D \zeta^2 - 8) (\zeta ( g_D \zeta -3) + 24 ) }{ \zeta^3 ( g_D \zeta^2  - 24 )^2 }\ .
\label{eq:dgdzeta}
\]
From Eq.~(\ref{eq:chi}) and Eq.~(\ref{eq:dgdzeta}), 
the singular behaviors may be observed at 
\[
\zeta^{\pm}_1 = \pm \frac{2\sqrt{2}}{\sqrt{ g_D }}\ , \ \ \ 
\zeta^{\pm}_2 = \frac{1}{ 2g_D } \left( 3 \pm \sqrt{ 9 - 96 g_D } \right)\ . 
\label{eq:singularzetas}
\]

When $g_D < 0$, one should choose $\zeta^-_2$ as the root. 
Since at $\zeta =  \zeta^-_2$, 
\[
\frac{\text{d} \chi}{\text{d} \zeta} \biggl{|}_{\zeta = \zeta^-_2}
= \frac{9 \left( -3 + \sqrt{9 - 96 g_D} \right)^3 \left( -3 + \sqrt{9 - 96 g_D}  + 32 g_D \right)^2 }{ 2048 g^3_D \left(  -9  + \sqrt{9 - 96 g_D}  + 64 g_D \right)^2  } \ne 0\ , 
\]
one obtains at the leading order 
\[
\chi (\zeta) - \chi (\zeta^-_2) 
\sim (\zeta - \zeta^-_2) \ . 
\label{eq:chiminuschic}
\]
Evaluating Eq.~(\ref{eq:dgdzeta}) around $\zeta^-_2$, one obtains at the leading order 
\[
\frac{\text{d} g}{\text{d} \zeta} \sim (\zeta - \zeta^-_2)\ , 
\label{eq:dgdzeta2}
\]
and therefore
\[
g (\zeta) - g(\zeta^-_2) \sim (\zeta - \zeta^-_2)^2 \ .  
\label{eq:gminusgc}
\]
From Eq.~(\ref{eq:chiminuschic}) and Eq.~(\ref{eq:gminusgc}), the susceptibility behaves around $g_c = g(\zeta^-_2)$ as
\[
\chi \sim  (g_c - g)^{-\gamma}\ , \ \ \ \text{with}\ \ \ \gamma = - \frac{1}{2}\ . 
\label{eq:chicrit}
\]
Therefore, one enters the pure-gravity phase approaching the critical line, $g_c(g_D)$ with $g_D<0$. 

Let us discuss what happens for $g_D \ge 0$. In the regime, $0 \le g_D < 9/128$, the singular behavior originates solely with $\zeta^-_2$,  
meaning that when $-\infty < g_D <  9/128$, one stays in the pure-gravity phase characterized by $\gamma = -1/2$.

However, if one reaches $g_D = g^*_D := 9/128$, the two roots coincide, i.e. $\zeta^+_1 = \zeta^-_2 = \zeta_{\ast}$, 
which yields
\[
g_D^* = \frac{9}{128}\ , \ \ \ \zeta_* = a^2_* = \frac{32}{3}\ , \ \ \ g_{\ast} = g_c(g_D^*) = \frac{3}{64}\ .
\label{eq:criticalcouplings}
\]
At this critical point, the denominator of $\chi$ diverges, which indicates the positive value of $\gamma$. 
Around this critical point, the susceptibility (\ref{eq:chi}) behaves as 
\[
\chi 
\sim \frac{1}{\zeta - \zeta_{\ast}} \ .
\label{eq:chiast}
\]
From Eq.~(\ref{eq:gofzeta}), one obtains
\[
g-g_{\ast} \sim  \left( \zeta - \zeta_{\ast} \right)^3 \ ,
\label{eq:gast}
\] 
and hence one obtains
\[
\chi \sim (g_{\ast} -g)^{-1/3} \ ,
\label{eq:gammawormhole}
\]
which means that $\gamma = + 1/3$. Approaching this critical point (\ref{eq:criticalcouplings}), one reaches the wormhole-dominant phase.

For $9/128 < g_D < \infty$, one should chose $\zeta^+_1$. 
In this region, the susceptibility (\ref{eq:chi}) behaves as
\[
\chi \sim \frac{1}{\zeta - \zeta^+_1}\ . 
\label{eq:chibp}  
\]
Evaluating Eq.~(\ref{eq:dgdzeta}) around $\zeta^+_1$, one finds
\[
\frac{\text{d}g}{\text{d}\zeta} \sim \zeta - \zeta^+_1\ .    
\label{eq:dgdzeta1}
\]
If setting $g_c := g(\zeta^+_1)$, Eq.~(\ref{eq:dgdzeta1}) yields
\[
g_c - g \sim (\zeta - \zeta^+_1)^2\ . 
\label{eq:gcminusgbp}
\]
Using Eq.~(\ref{eq:chibp}) and Eq.~(\ref{eq:gcminusgbp}), 
\[
\chi \sim (g_c-g)^{-1/2} \ ,
\label{eq:gammabp}
\]
which shows that $\gamma = + 1/2$. 
Therefore, in the region, $9/128 < g_D < \infty$, the branched-polymer phase appears.

We next wish to calculate the free energy near the wormhole-dominant phase. 
Using Eq.~(\ref{eq:gofzeta}) and setting $g_D = g_D^*= 9/128$, we expand the free energy (\ref{eq:free_energy2}) around $\zeta_{\ast} = 32/3$:
\[
F \cong 
\frac{83}{72} - \frac{1}{2} \log\frac83
+ \frac{3^2}{2^{16}} (\zeta_* - \zeta)^3 + \frac{3^4}{2^{23}} (\zeta_* - \zeta)^4 
- \frac{3^5}{5\cdot2^{28}} (\zeta_* - \zeta)^5 + \cdots \ .
\label{eq:expandfaroundzetaast} 
\]
On the other hand, if we expand $g$ around $\zeta_*$ 
\[
-\frac{16}{9}g \cong
- \frac{1}{12}  
+ \frac{3^2}{2^{16}} (\zeta_* - \zeta)^3 + \frac{3^4}{2^{23}} (\zeta_* - \zeta)^4 
+ \frac{3^6}{2^{28}}(\zeta_* - \zeta)^5 + \cdots\ ,
\label{eq:expandgaroundzetaast}
\]
which means 
\[
g_* - g \cong 
\frac{3^4}{2^{20}} (\zeta_* - \zeta)^3 + \frac{3^6}{2^{27}} (\zeta_* - \zeta)^4
+ \frac{3^8}{2^{32}}(\zeta_* - \zeta)^5 + \cdots 
\ , 
\label{eq:expandgaroundzetaast2}
\]
where $g_* = 3/64$. 
Therefore, the free energy behaves near the critical point of the wormhole-dominant phase as
\[
F \cong  \frac{83}{72} - \frac{1}{2} \log\frac83 
+ \frac{16}{9} (g_{\ast} - g) 
- \frac{2^{\frac{28}{3}}}{5\cdot3^{\frac53}} (g_*-g)^{5/3} + \cdots\ . 
\label{eq:freeenergywormhole}
\]
This is consistent with the fact that $\gamma=+1/3$ at the wormhole-dominant phase. 

Likewise, in the case of $W(z)=z^4$, Eq.~\eqref{eq:tenthorderequation} leads to 
\[
\frac{\text{d}g}{\text{d}\zeta}=\frac{1}{48\zeta^3}(\zeta-8)(9g_D\zeta^4-64)\ .
\]
The two solutions to it 
\[
\zeta_1 = 8\ , \ \ \
\zeta_2^4 = \frac{64}{9g_D}\ ,  
\]
agree when 
\[
g_D=g_D^*=\frac{1}{2^6\cdot 3^2},
\]
and then $g_*=1/18$. These results exactly coincide with the ones in Ref.~\cite{Klebanov:1994kv}. 
Following the derivation from Eq.~\eqref{eq:expandfaroundzetaast} to 
Eq.~\eqref{eq:freeenergywormhole}, we get 
\[
F \cong  \frac{19}{18} - \frac{1}{2} \log2
+  (g_{\ast} - g) 
- \frac{3^{\frac{10}{3}}}{5\cdot 2^{\frac23}} (g_*-g)^{5/3} + \cdots\ . 
\label{eq:freeenergywormhole2}
\] 
We can check that the coefficient of the universal term of ${\cal O}((g_*-g)^{5/3})$ 
agrees with the one in Ref.~\cite{Klebanov:1994kv} after matching the conventions carefully. 

Finally in order to confirm that the two approaches proposed in Refs.~\cite{Das:1989fq,Klebanov:1994kv} in fact give the same critical point\footnote
{Due to universality, we indeed anticipate that details of $W(z)$ does not affect 
universal behavior, as mentioned briefly in \cite{Klebanov:1994pv}.}, 
let us apply the latter one in Eq.~\eqref{eq:KH} to the model considered in the former, 
namely the case $V(z)=\frac12z^2-\frac g4z^4$, $W(z)=z^2$. 

Let us essentially follow the discussion in Sec.~\ref{sec:introductionofwormholes}, and formally consider the integral: 
\[
Z_N(g, g_D) 
&= \int^{\infty}_{-\infty} \text{d}\nu\, \rho_G(\nu)\,
Z_{\nu}(g)\ ,\notag \\
Z_{\nu}(g)&=\frac{1}{\Omega_N}
\int D\phi\, 
e^{-N \text{tr} \left( \frac12(1+2\nu)\phi^2-\frac g4\phi^4 \right)}
\notag \\
&=(1+2\nu)^{-\frac{N^2}{2}}
\frac{1}{\Omega_N}
\int D\phi\, 
e^{-N \text{tr} \left( \frac12\phi^2-\frac{g}{4(1+2\nu)^2}\phi^4 \right)}
=(1+2\nu)^{-\frac{N^2}{2}}Z_{\nu=0}\left(\frac{g}{(1+2\nu)^2}\right)\ ,
\label{eq:HKforW4}
\]
where $\rho_G (\nu)$ is the probability density function of the Gaussian distribution defined by Eq.~(\ref{eq:gaussian}). 

The equation corresponding to Eq.~(\ref{eq:KH2}) yields
\[
e^{- \mathcal{F} (g,g_D)} =  
\int^{\infty}_{- \infty}\text{d} \nu\ \rho_G(\nu)\ e^{-  \mathcal{F}( g/(1+2\nu)^2 , 0) - \frac{N^2}{4}\log\left(1+2\nu\right)^2}\ , \ \ \ \text{with}\ \ \ \mathcal{F} (g,g_D) :=  - \log Z_N (g,g_D)\ . 
\label{eq:KH3}
\]  
If we formally change the variable $\nu$ to $t$, 
\[
\frac{g}{(1+2\nu)^2} = \frac{1}{12} (1 - \varepsilon^2 t) \ , 
\label{nutot2}
\]
we then obtain the function $f(t)$ analogous to the one defined in Eq.~(\ref{eq:regularization2}) for $t>0$ and $\nu>0$:
\[
f(t) &\cong 
 N^2 \biggl(
f_0 + f_1\, \varepsilon^2 t  + f_2\, (\varepsilon^2 t)^2 + f_{5/2}\, (\varepsilon^2 t)^{\frac52}+{\cal O}\left((\varepsilon^2 t)^3\right) 
\biggl) + \mathcal{O}(N^0)\ , 
\label{eq:ft3}
\]
where 
\[
f_0 &= \frac{25}{24}+\frac14\log(3g)+\frac{12g-4\sqrt{3g}+1}{8g_D}\ , \label{eq:coefficient0} \\
f_1 &= \frac13+\frac{6g-\sqrt{3g}}{4g_D}\ , \ \ \ 
f_2 = \frac{3\left(8g-\sqrt{3g}\right)}{16g_D}\ , \ \ \ 
f_{5/2} = \frac{4}{15}\ . 
\label{eq:coefficients2}
\]
The prescription in Ref.~\cite{Klebanov:1994kv} amounts to requiring 
that both $f_1$ and $f_2$ vanish, which yields 
\[
g_D^*=\frac{9}{128}\ , \ \ \ g_*=\frac{3}{64}\ .
\]
This is the same as the one in Eq.~\eqref{eq:criticalcouplings}.

\section{Continuum limit}
\label{sec:continuumlimit}

In the context of the model of random surface, considering the resolvent is to introduce a single marked boundary to random surfaces where $z$ is interpreted as the (dimensionless) boundary cosmological constant. 
Therefore, the resolvent can be interpreted as the marked disk amplitude.      
Thus in this section let us take the continuum limit of the resolvent. In particular, we focus on the wormhole-dominant phase.  

In the case of $V(z)=\frac12z^2-\frac g4z^4$ and $W(z)=z^2$, setting $g_D = g_D^*$ first, we then tune $g$ to $g_*$ and $z$ to $z_*$, through the use of the following parametrization:
\[
g=g_* e^{- (\varepsilon^2 \Lambda)^{3/2}} \ , \ \ \ 
z=a_* e^{\varepsilon Z}\ , \ \ \  a^2=a^2_* e^{-\varepsilon C}\ , 
\label{eq:parametrization}
\]
where $g_{\ast} = 3/64$, and $a_* = \sqrt{\zeta_*} =  \sqrt{32/3}$. 
We have introduced the lattice spacing $\varepsilon$, and the renormalized bulk and boundary cosmological constants, $\Lambda$ and $Z$, whose mass dimensions are $2$ and $1$, respectively. 
The dimensionful constant $C$ will be determined as a function of $\Lambda$ in due course. 
The parametrization (\ref{eq:parametrization}) has been chosen in such a way as to be consistent with Eq.~(\ref{eq:gast}).  

Plugging the parametrization (\ref{eq:parametrization}) into Eq.~(\ref{eq:eighthorderequation}), 
one can determine $C$:
\[
C= 2^{-1/3} \sqrt{\Lambda}\ . 
\label{eq:c}
\]
With the scaling above, the resolvent becomes
\[
R_0(z) \cong 
\frac{1}{\sqrt{6}} 
- \sqrt{\frac{3}{2}}Z \varepsilon 
+ \frac{4}{\sqrt{3}} \varepsilon^{3/2} W_{\Lambda}(Z) + \mathcal{O}(\varepsilon^2)\ ,
\label{eq:expandw}
\]
where the continuum limit of the resolvent is
\[
W_{\Lambda}(Z) = 
\left( Z - \frac{1}{2}\sqrt{ 2^{-8/3} \Lambda} \right)
\sqrt{
Z+
 \sqrt{2^{-8/3} \Lambda} 
}\ . 
\label{eq:continuumw}
\]
The inverse Laplace transform of Eq.~(\ref{eq:continuumw}) is consistent with the loop correlator obtained in Ref.~\cite{Barbon:1995dx}. 

When $Z \gg \sqrt{\Lambda}$, the one-dimensional boundary would shrink to a point, 
and one can read off the string susceptibility: 
\[
W_{\Lambda}(Z) \cong  Z^{3/2} - \frac{3}{2^{17/3} Z^{1/2}} \Lambda + \mathcal{O}(Z^{-3/2}) \sim \Lambda \sim (g_{\ast}-g)^{1-1/3}  \ . 
\label{eq:wforlargez}
\]
Therefore, we obtain $\gamma=+1/3$, as expected. 

Similarly, in the case of $V(z)=\frac12z^2-\frac g4z^4$ and $W(z)=z^4$, 
we adopt the same scaling as in Eq.~\eqref{eq:parametrization} with 
$g_*=1/18$ and $g_D^*=1/3^2\cdot2^6$, and find 
\[
C= 2^{-2/3} \sqrt{\Lambda}\ . 
\] 
Then we get 
\[
R_0(z) \cong 
\frac{\sqrt{2}}{3}
- \sqrt{2}Z \varepsilon 
+ \frac{8}{3} \varepsilon^{3/2} W_{\Lambda}(Z) + \mathcal{O}(\varepsilon^2)\ ,
\]
where 
\[
W_{\Lambda}(Z) = 
\left( Z - \frac{1}{2}\sqrt{ 2^{-10/3} \Lambda} \right)
\sqrt{
Z+
 \sqrt{2^{-10/3} \Lambda} 
}\ . 
\]
We have thus confirmed the universality explicitly, and found that 
the disk amplitude in the unconventional Liouville gravity coincides with 
the conventional one.\footnote
{It would be intriguing to derive this using Eq.~\eqref{eq:Laplace} 
by, for example, introducing coupling constants.}  

From the Liouville theory side, this fact can be understood as follows: 
In the conventional branch, we start from the action:  
\[
S_L=\frac{1}{4\pi}\int_{\cal M} \text{d}^2x\sqrt{g}\left(g^{ab}\partial_a\phi\partial_b\phi
+ QR\phi+4\pi\mu e^{2b\phi}\right)
+\int_{\partial{\cal M}} \text{d}\tau\sqrt{\gamma}\left(\frac{Q\phi}{2\pi}K+\mu_B e^{b\phi}\right)\ ,
\label{eq:SL}
\]
where $\mu$ and $\mu_B$ are the bulk and boundary cosmological constant respectively, The Liouville central charge $c_L$, the background charge $Q$, and the parameter $b$ are related by 
\[
c_L=1+6Q^2\ , \ \ \ Q=b+\frac1b\ .
\]
In our pure gravity case, $Q=5/\sqrt 6$ and $b$ has two solutions: 
$b=\sqrt{2/3}, \sqrt{3/2}$, where we take the former in the conventional branch.  
In the unconventional branch, we instead take the latter and hence 
$b$ is replaced by $1/b$ in the action \eqref{eq:SL}.

The disk one-point function with boundary length $\ell$ for the operator 
$V_{\alpha}(z)=e^{2\alpha\phi}$ is given by Refs.~\cite{Fateev:2000ik,Teschner:2000md}:
\[
W_{\alpha}(\ell)\equiv\vev{V_\alpha(z)}=\frac2b(\pi\mu\gamma(b^2))^{\frac{Q-2\alpha}{2b}}
\frac{\Gamma(2b\alpha-b^2)}{\Gamma\left(1+\frac{1}{b^2}-\frac{2\alpha}{b}\right)}
K_{\frac{Q-2\alpha}{b}}(\kappa \ell)\ , 
\label{eq:FZZ}
\]
where 
\[
\kappa=\sqrt{\frac{\mu}{\sin\left(\pi b^2\right)}}\ .
\]
In the conventional branch, $V_b(z)$ is exactly the holomorphic part of the Liouville potential term in the action \eqref{eq:SL} and hence one can derive the disk amplitude 
without operator insertion by integrating $W_{\alpha}(\ell)$ 
in Eq.~\eqref{eq:FZZ} with respect to $\mu$ as 
\[
W(\ell)=-\int \text{d} \mu\,W_b(\ell)
=\frac4b(\pi\mu\gamma(b^2))^{\frac{1-b^2}{2b^2}}
\frac{\Gamma(b^2)}{\Gamma\left(-1+\frac{1}{b^2}\right)}\frac{\mu}{\kappa \ell}
K_{\frac{1}{b^2}}(\kappa \ell)\ .
\]
Since the disk amplitude in the matrix model is expressed in terms of 
the bulk and boundary cosmological constant as in Eq.~\eqref{eq:continuumw}, we make the Laplace transformation 
with respect to $\ell$ to obtain 
\[
\int_0^{\infty} \text{d}\ell\ e^{-\mu_B \ell}W(\ell)
=&\frac2b(\pi\mu\gamma(b^2))^{\frac{1-b^2}{2b^2}}
\sqrt{\mu \sin\left(\pi b^2\right)}
\frac{\Gamma(b^2)}{\Gamma\left(-1+\frac{1}{b^2}\right)}
\Gamma\left(\frac{1}{b^2}\right)\Gamma\left(-\frac{1}{b^2}\right) \notag \\
&\times
\left(
\left(\frac{\mu_B'}{\sqrt{\mu}}+\sqrt{\frac{\mu_B'^2}{\mu}-1}\right)^{\frac{1}{b^2}}
+\left(\frac{\mu_B'}{\sqrt{\mu}}-\sqrt{\frac{\mu_B'^2}{\mu}-1}\right)^{\frac{1}{b^2}}
\right)\ ,
\label{eq:Liouvilledisk}
\]
where $\mu_B'=\sqrt{\sin\left(\pi b^2\right)}\mu_B$. Recalling $b=\sqrt{2/3}$, 
we find that the above equation has the same functional form as in Eq.~\eqref{eq:continuumw} as 
\[
\int_0^{\infty} \text{d}\ell\ e^{-\mu_B \ell}W(\ell)
\propto \left(\mu_B'-\frac{\sqrt{\mu}}{2}\right)\sqrt{\mu_B'+\sqrt{\mu}}
\]
under the suitable identification.  
On the other hand, if we try to follow the same procedure in the unconventional branch where the action contains $V_{\frac1b}(z)$, we would first consider 
$W_{\frac{1}{b}}(\ell)$, integrate it with respect to $\mu$, and make the Laplace transformation. However, Eq.~\eqref{eq:FZZ} tells us that $W_{b}(\ell)$ 
and $W_{\frac1b}(\ell)$ are the same as a function of $\ell$ only with difference in the numerical factors. Hence even in the unconventional branch, we essentially get the same disk amplitude in Eq.~\eqref{eq:Liouvilledisk}. Note that this observation can be regarded as another support for the claim that the double-trace matrix model provides nonperturbative formulation of the Liouville theory in the unconvetional branch.

\subsection{Relevant deformation}
\label{sec:relecant}
In the previous study $g_D$ has been fixed on its critical value. 
However, it would be interesting to observe what happens if we take the continuum limit away from it.
We thereby introduce a new coupling constant associated with the touching interaction in a continuum theory.

Let us approach the critical point for the wormhole-dominant phase in a manner different from Eq.~(\ref{eq:parametrization}):  
\[
g=g_* e^{- (\varepsilon^2 \Lambda)^{3/2}} \ , \ \ \ 
z=a_* e^{\varepsilon Z}\ , \ \ \ 
a^2=a^2_* e^{-\varepsilon \widetilde{C}}\ , \ \ \ 
g_D = g_D^* e^{- \varepsilon^n \Theta}\ , 
\label{eq:parametrization2}
\]
where $n$, a positive constant, and $\widetilde{C}$ will be determined in a consistent way. 
Here $\Theta$ is the renormalized coupling for the wormhole interaction; 
the positive (negative) $\Theta$ means that one approaches the critical point $g_D^*$ from the pure-gravity phase (the branched polymer phase). 

As before, plugging the parametrization (\ref{eq:parametrization2}) into Eq.~(\ref{eq:eighthorderequation}), 
one can determine $\widetilde{C}$:
\[
\widetilde{C}= 2^{-1/3} \left( \Lambda^{3/2} + \Theta \right)^{1/3} \ . 
\label{eq:c2}
\]
In order to get the non-trivial scaling, we need to set $n=3$\ \footnote{
Since we know the coupling $g_D$ is associated with the microscopic wormhole that connects two surfaces, 
it might be better to use $(\varepsilon^4 \Delta)^{3/4}$ instead of $\varepsilon^3 \Theta$ where the mass dimension of $\Delta$ is $4$. 
However, it is not an essential issue which one, $\Delta$ or $\Theta$, is used because it is just a redefinition of the renormalized coupling constant. 
}. 
With the scaling above with $n=3$, the resolvent becomes
\[
w(z) \cong 
\frac{1}{\sqrt{6}} 
- \sqrt{\frac{3}{2}}Z \varepsilon 
+ \frac{4}{\sqrt{3}} \varepsilon^{3/2} W_{\Lambda_{\text{eff}}}(Z) + \mathcal{O}(\varepsilon^2)\ .
\label{eq:expandw2}
\]
Here the continuum limit of the resolvent is
\[
W_{\Lambda_{\text{eff}}}(Z) = 
\left( Z - \frac{1}{2}\sqrt{ 2^{-8/3} \Lambda_{\text{eff}}} \right)
\sqrt{
Z+
 \sqrt{2^{-8/3} \Lambda_{\text{eff}}} 
}\ ,
\label{eq:continuumw2}
\]
where the effective bulk cosmological constant is given by
\[
\Lambda_{\text{eff}} =
\Lambda \left( 1 + s \right)^{2/3}\ , \ \ \ \text{with} \ \ \
s = \Theta / \Lambda^{3/2}\ .  
\label{eq:effectivelambda}
\]
The coupling for the wormhole interaction is essentially absorbed, and deforms the bulk cosmological constant effectively\footnote{
This cannot be anticipated just from dimensional analysis.  
}. 

Even if the bulk cosmological constant $\Lambda$ is positive, the effective bulk cosmological constant $\Lambda_{\text{eff}}$ can be zero. 
When approaching the critical point from the branched polymer phase, i.e. $\Theta <0$, 
in order for $\Lambda_{\text{eff}}$ to be real, the allowed range of the dimensionless parameter $s$ is
\[
-1 \le s < 0\ . 
\label{eq:rangeoft}
\] 
When $s =-1$, the effective cosmological constant vanishes. 
Therefore, if the wormhole coupling comes into balance with the original bulk cosmological constant, i.e. $\Theta = - \Lambda^{3/2}$, 
the wormhole effects force the effective bulk cosmological constant to be zero.  
This interplay between the bulk cosmological constant and the wormhole effect 
is reminiscent of the mechanism proposed by Coleman \cite{Coleman:1988tj}, 
but there are several differences: for example, our free energy does not seem to have a sharp peak at a small cosmological constant. Their relationship would deserve further study.

\section{Nonperturbative effect}
\label{sec:NE}

In this section we derive nonperturbative effect in the scaling limit \eqref{eq:parametrization} of our matrix model following the derivation 
in Ref.~\cite{Hanada:2004im}. 
Since it is given as the integration of the disk function,  
which we have shown to match the one in the pure-gravity phase, 
it seems to be trivial that 
the nonperturbative effect also coincides. 
However, the scaling limit \eqref{eq:parametrization} itself is different from the model without 
double-trace term. 
Moreover, as shown in Ref.~\cite{Klebanov:1994kv}, 
the perturbative expansion of the free energy is different from the standard one 
as a function of the bulk renormalized cosmological constant 
since it is obtained by the Laplace transformation \eqref{eq:Laplace}, 
and it is therefore nontrivial whether nonperturbative effects are also the same 
for conventional and unconventional Liouville theory from the point of view of resurgence. 

We begin with the partition function in terms of eigenvalues in Eq.~\eqref{eq:dtmm2} 
and by calling one of them $x$, it is rewritten as
\[
Z_N(g,g_D) 
=&\int\prod_{i=1}^{N} \text{d} \lambda_i\,\Delta^{(N)}(\lambda)^2
e^{-N\sum_{i=1}^NV(\lambda_i)
+\frac{g_D}{2}\left(\sum_{i=1}^NW(\lambda_i)\right)^2} \notag \\
=&N\int \text{d} x\int\prod_{i=1}^{N-1}d\lambda'_i\,\Delta^{(N-1)}(\lambda')^2
\prod_{i=1}^{N-1}(x-\lambda'_i)^2 \notag \\
&\times e^{-N\sum_{i=1}^{N-1}V(\lambda'_i)+g_DW(x)\sum_{i=1}^{N-1}V(\lambda'_i)
+\frac{g_D}{2}\left(\sum_{i=1}^{N-1}W(\lambda'_i)\right)^2}
e^{-NV(x)+\frac{g_D}{2}W(x)^2} \notag \\
=&N\int \text{d} x\,\vev{\det\left(x-\phi'\right)^2
e^{g_DW(x)\text{tr}\,W(\phi')}}'_{N-1}Z'_{N-1}e^{-NV(x)+\frac{g_D}{2}W(x)^2}\ , 
\]
where 
\[
Z_{N-1}'=&\frac{1}{\Omega_{N-1}}\int D\phi'\,e^{-N\text{tr}\,V(\phi')
+\frac{g_D}{2}\left(\text{tr}\,W(\phi')\right)^2}\ , \notag\\
\vev{{\cal O}}'_{N-1}
=&\frac{1}{Z'_{N-1}\Omega_{N-1}}\int D\phi'\,{\cal O}\,
e^{-N\text{tr}\,V(\phi')+\frac{g_D}{2}\left(\text{tr}\,W(\phi')\right)^2
}\ ,
\]
are quantities in the rank $N-1$ one-matrix model with the factor in front of 
tr replaced with $N$ which is why we have put the prime. Hence we get 
\[
\frac{Z_N}{Z'_{N-1}}=N\int \text{d} x\,e^{-NV_1(x)}\ ,
\]
where 
\[
V_1(x)&:=V(x)+\frac{g_D}{2N}W(x)^2-\frac{1}{N}\log
\vev{\det\left(x-\phi'\right)^2
e^{g_DW(x)\text{tr}\,W(\phi')}}'_{N-1} \notag \\
&=V(x)+\frac{g_D}{2N}W(x)^2-\frac{1}{N}\log
\vev{e^{2\, \text{Re}\, \text{tr}\,\log\left(x-\phi'\right)
+g_DW(x)\text{tr}\,W(\phi')}}'_{N-1} \notag \\
&=V(x)+\frac{g_D}{2N}W(x)^2-\frac{1}{N}\sum_{n=1}^{\infty}\frac{1}{n!}
\vev{
\left(2\, \text{Re}\, \text{tr}\,\log\left(x-\phi'\right)
+g_DW(x)\text{tr}\,W(\phi')\right)^n
}'_{N-1,c}\ . 
\]
In this equation $c$ denotes the sum of connected diagrams. 
Thus in the large-$N$ limit, $V_1(x)$ becomes 
\[
V_1^{(0)}(x)=V(x)-2\text{Re}\, \vev{\frac1N\text{tr}\,\log(x-\phi)}_0-g_DW(x)W_0\ , 
\]
in which the second term on the right-hand side can be evaluated 
by using Eq.~\eqref{eq:resolvent} as 
\[
\text{Re}\, \vev{\frac1N\text{tr}\,\log(x-\phi)}_0
=\int_{L}^x\text{Re}\, R_0(x)+\log L\ , \ \ \ (L\rightarrow\infty)\ .
\]
Now the saddle point equation reads 
\[
0=&V_1^{(0)\prime}(x)=V'(x)-2\text{Re}\, R_0(x)-g_DW_0W'(x) \notag \\
=&\text{Re}\, \sqrt{\widetilde{V}'(x)^2-4Q_0(x)}\ ,
\label{eq:speforV1}
\]
where we have used the fact that $R_0(x)$ satisfies Eq.~\eqref{eq:loopeq}. 

As a concrete example, let us consider the case $V(z)=\frac12z^2-\frac g4z^4$, $W(z)=z^2$, where the resolvent takes the form in Eq.~\eqref{eq:R0forW4}. 
Thus straightforward calculation shows that for $x\geq a$,\footnote
{As shown in Ref.~\cite{Hanada:2004im}, $V_1^{(0)}$ becomes flat for $x\in [-a,a]$, 
where $\text{Re}\sqrt{x^2-a^2}=0$.} 
\[
V_1^{(0)}(x)=\frac{g}{8}\left(a^2(a^2-4b^2)\arccosh\frac xa-x\sqrt{x^2-a^2}
\left(2x^2-a^2-4b^2\right)+\frac14a^2(a^2+8b^2)\right)+2\log\frac a2\ .
\]
In particular, 
\[
V_1^{(0)\prime}(x)=-g(x^2-b^2)\sqrt{x^2-a^2}\ ,
\]
which in fact agrees with Eq.~\eqref{eq:speforV1}, and 
\[
V_1^{(0)}(x)-V_1^{(0)}(a)=-2\arccosh\frac xa-\frac{1}{4}gx\sqrt{x^2-a^2}\left(x^2-a^2-\frac{8}{ga^2}\right)\ ,
\label{eq:V1}
\]
where we have used Eq.~\eqref{eq:bbya}. Therefore it is apparent that for $x\geq a$, 
$V_1^{(0)\prime}=0\, \Leftrightarrow\, x=a, b$. Thus we obtain nonperturbative effect in the large-$N$ limit as
\[
V_1^{(0)}(b)-V_1^{(0)}(a)=-2\arccosh\frac ba-\frac g4b\sqrt{b^2-a^2}
\left(b^2-a^2-\frac{8}{ga^2}\right)\ . 
\]
Plugging Eq.~\eqref{eq:bbya} and taking the limit in Eq.~\eqref{eq:parametrization}, 
this becomes 
\[
V_1^{(0)}(b)-V_1^{(0)}(a)=\frac{2^{\frac53}3^{\frac12}}{5}\varepsilon^{\frac52} \Lambda^{\frac54}
+{\cal O}\left(\varepsilon^{\frac72}\right). 
\]
This would be prediction of tension of a brane in the unconventional branch of Liouville theory. It is intriguing to examine whether it is also derived by using Eq.~\eqref{eq:Laplace}, or applying resurgence to the perturbative series 
of the free energy given in Ref.~\cite{Klebanov:1994kv}.

\section{Conclusions and Discussions}
\label{sec:discussion}

We have studied the double-trace matrix model in the large-$N$ limit, focusing in particular on the wormhole-dominant phase. 

We have first calculated the continuum limit of the resolvent that is the marked disk amplitude. 
The resulting continuum disk amplitude is the same as that of the pure $2$D quantum gravity, 
even though the string susceptibility is not $\gamma = -1/2$ but $\gamma = + 1/3$.  
 
Additionally, we have shown that one can introduce the renormalized coupling for the double-trace interaction in such a way as to survive in the continuum limit. 
The newly introduced coupling is absorbed into the renormalized bulk cosmological constant, and therefore can effectively change the value of the bulk cosmological constant. 
The effective bulk cosmological constant can in principle vanish by the fine tuning of the newly-introduced coupling, even if the original bulk cosmological constant is positive. 
This phenomenon originates from the proliferation of microscopic wormholes.   

The relation between the sum over wormholes and the bulk cosmological constant was discussed by Coleman \cite{Coleman:1988tj}, 
suggesting that the bulk cosmological constant effectively tends to zero, which is induced by the sum over non-local operators, i.e. wormholes. 
This Coleman mechanism is a realization of the self-organization through wormholes, and therefore does not require any fine-tuning of the couplings.   
Our result may not have direct relation to the Coleman mechanism since making the bulk cosmological constant vanish requires the fine-tuning of the renormalized coupling for the double-trace interaction, but we hope that it reveals a new aspect 
between the cosmological constant and wormholes.

\begin{figure}[h]
\centering
\begin{minipage}[b]{0.49\columnwidth}
    \centering
    \includegraphics[width=0.9\columnwidth]{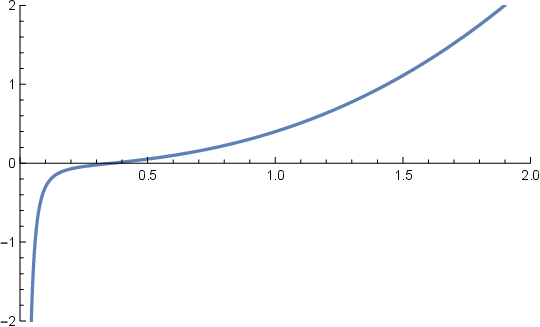}
    \caption{A plot for $\mathcal{F} (t)$. }
    \label{fig:a}
\end{minipage}
\begin{minipage}[b]{0.49\columnwidth}
    \centering
    \includegraphics[width=0.9\columnwidth]{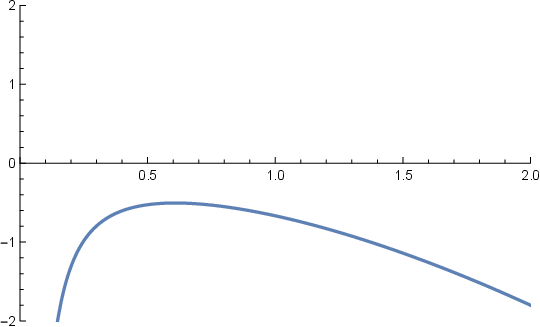}
    \caption{A plot for $\overline{ \mathcal{F}} (\bar t)$.}
    \label{fig:b}
\end{minipage}
\end{figure}

Here is a speculation: In Ref.~\cite{Klebanov:1994kv}, 
the minus of the free energies in the pure-gravity phase and the wormhole-dominant phase which are related by Eq.~\eqref{eq:Laplace} 
are explicitly given as 
\[
&  \mathcal{F} (t) 
\cong  \frac25t^{\frac52} + \frac{1}{48}\log t - \frac{7}{8640}t^{-\frac52}+{\cal O}\left(t^{-10}\right)\ ,\label{eq:puregravityF} \\ 
&  \overline{ \mathcal{F}} (\bar t) 
\cong -\frac35{\bar t}^{\frac53}+\frac{13}{72}\log \bar t
-\frac{257}{3840}{\bar t}^{-\frac53}+{\cal O}\left({\bar t}^{-\frac{10}{3}}\right)\ ,
\label{eq:wormholeF}
\]
If we pick up to the third term in each, they look as in Fig.~\ref{fig:a} and Fig.~\ref{fig:b}. 

We immediately find that crucial difference in these plots is attributed to the sign of the sphere free energy. 
Hence if the Coleman mechanism works only in the wormhole-dominant phase, there is a possibility that the sign of the free energy may play a key role in realization of the Coleman mechanism. 
Of course these plots are far from complete, because 
in the small $t$ region, the expansion in Eq.~\eqref{eq:puregravityF} 
and Eq.~\eqref{eq:wormholeF} cannot be justified and, in particular, we have to take account of the nonperturbative effect. 
However, it would be worth analyzing whether it is the case.

\section*{Acknowledgement}
We would like to thank Shinji Hirano and Taro Kimura   
for discussions and encouragements.   
The work of T.K. and Y.S.  was partially supported by JSPS KAKENHI Grant Number 22K03606 and 19K14705, respectively.

\appendix

\section{Schwinger-Dyson equation of the resolvent}
\label{app:SD}
In this appendix we derive the Schwinger-Dyson equation for the resolvent 
in the one-matrix model with the double-trace term in Eq.~\eqref{eq:dtmm}. 
Starting from the identity, 
\[
0=\int D\phi \frac{\delta}{\delta\phi^a}\frac{1}{N^2}\text{tr}\left(\frac{1}{z-\phi}\right)
e^{-NS}\ ,
\]
where $\phi=\sum_a\phi^at^a$ with $t^a$ the generator of U($N$), and 
\[
S=\text{tr}\, V\left(\phi\right) - \frac{g_D}{2N} \left( \text{tr}\, W\left(\phi\right)\right)^2\ , 
\] 
we get 
\[
0=\int D\phi\Biggl[&\left(\frac1N\text{tr}\,\frac{1}{z-\phi}\right)^2
+\frac1N\text{tr}\,\frac{V'(z)-V'(\phi)}{z-\phi}-V'(z)\frac1N\text{tr}\,\frac{1}{z-\phi}
\notag \\
&-g_D\left(\frac1N\text{tr}\,W(\phi)\right)
\left(\frac1N\text{tr}\,\frac{W'(z)-W'(\phi)}{z-\phi}-W'(z)\frac1N\text{tr}\,\frac{1}{z-\phi}\right)\Biggr]\ ,
\]
which yields 
\[
\vev{R(z)^2}+\vev{f_V(z)}-V'(z)\vev{R(z)}-g_D\left(\vev{Wf_W(z)}-W'(z)\vev{WR(z)}\right)=0\ ,
\label{eq:SDforR}
\]
where $\langle \cdot \rangle$ denotes the expectation value in the model \eqref{eq:dtmm}, 
and 
\[
&R(z):=\frac1N\text{tr}\,\frac{1}{z-\phi}\ , \notag\\ 
&W:=\frac1N\text{tr}\,W(\phi)\ , \notag\\
&f_U(z):=\frac1N\text{tr}\,\frac{U'(z)-U'(\phi)}{z-\phi}\ , 
\text{ for a function } U(z) \ .
\] 
Introducing a $\phi$-dependence potential analogous to $\tilde V(z)$ defined in Eq.~\eqref{eq:W0} 
\[
\tilde{{\cal V}}(z):=V(z)-g_DWW(z),
\]
Eq.~\eqref{eq:SDforR} is expressed in terms of $\tilde{{\cal V}}(z)$ as  
\[
\vev{\left(R(z)-\tilde{{\cal V}}'(z)\right)R(z)}+\vev{f_{\tilde{\cal V}}(z)}=0.
\]
In the large-$N$ limit, we invoke the factorization to obtain the quadratic equation for the resolvent: 
\[
R_0(z)^2-\tilde{V}'(z)R_0(z)+\vev{f_{\tilde{V}}(z)}_0=0\ .
\]

\end{document}